\documentclass[conference]{IEEEtran}
\IEEEoverridecommandlockouts
\usepackage[table,xcdraw]{xcolor}

\usepackage{booktabs, makecell, multirow, tabularx}
\newcolumntype{L}{>{\raggedright\arraybackslash}X}
\usepackage{booktabs}
\usepackage[utf8]{inputenc}
\usepackage{xcolor}

\usepackage{times}
\usepackage{epsfig}
\usepackage{graphicx}
\usepackage{amsmath}
\usepackage{amssymb}
\usepackage[font=scriptsize,skip=0pt]{caption}
\usepackage{subfigure}
\usepackage{textcomp}
\usepackage{gensymb}
\usepackage[none]{hyphenat}
\usepackage{epstopdf}
\usepackage{balance}
\usepackage[hyphens]{url}
\usepackage[breaklinks=true]{hyperref}
\usepackage{breakcites}
\usepackage[numbers,sort&compress]{natbib}
\usepackage{fancyhdr}

\def\BibTeX{{\rm B\kern-.05em{\sc i\kern-.025em b}\kern-.08em
    T\kern-.1667em\lower.7ex\hbox{E}\kern-.125emX}}
    
\begin{document}
\title{Detection and Classification of Internal Faults in Power Transformers using Tree based Classifiers}
\author{
{\tt\small}
Samita Rani Pani\\
KIIT, Bhubaneswar, India\\
{\tt\small samita.panifel@kiit.ac.in}
\and
Pallav Kumar Bera\\
Syracuse University, USA \\
{\tt\small pkbera@syr.edu}
\and
Vajendra Kumar\\
IIT Roorkee, India \\
{\tt\small kumarvajendra@gmail.com}
}

\IEEEoverridecommandlockouts
\maketitle
\IEEEpubidadjcol

\begin{abstract}
This paper proposes a Decision Tree (DT) based detection and classification of internal faults in a power transformer. The faults are simulated in Power System Computer Aided Design (PSCAD)/ Electromagnetic Transients including DC (EMTDC) by varying the fault resistance, fault inception angle, and percentage of winding under fault. A series of features are extracted from the differential currents in phases a, b, and c belonging to the time, and frequency domains. Out of these, three features are selected to distinguish the internal faults from the magnetizing inrush and another three to classify faults in the primary and secondary of the transformer. DT, Random Forest (RF), and Gradient Boost (GB) classifiers are used to determine the fault types. The results show that DT detects faults with 100\% accuracy and the GB classifier performed the best among the three classifiers while classifying the internal faults.
\end{abstract}

\section{Introduction}

Power transformers are an integral part of any Power network. They are expensive and once damaged their repairs are time-consuming. Thus, their protection is vital for reliable and stable operation of the power system. Transformer-protective relays are often tested for their dependability, stability, and speed of operation under different operating conditions. The protective relays should operate in cases of faults and avoid tripping the circuit breakers when there is no fault. Many relays having different characteristics such as the current differential, Buchholz, Volts/Hz, over current, etc. protect the transformers in case of internal conditions (phase (ph)-phase faults, phase-g faults, inter-turn faults, over fluxing). Although differential protection has been the primary protection for many years, it suffers from traditional challenges like magnetizing inrush current, saturation of core, Current Transformer (CT) ratio mismatch, external fault with CT saturation, etc. Percentage differential relay with second harmonic restraint \cite{harmonic} which detects magnetizing inrush may fail because of the lower content of second harmonics in inrush currents of modern transformers \cite{modern_core} and presence of higher second harmonics in internal faults
with CT saturation and with distributed and series compensated lines \cite{tx3pro}.

Researchers have proposed different intelligent techniques to distinguish internal faults and magnetizing inrush and classify internal faults in power transformers in the past. Tripathy et al. used the Probabilistic Neural Network (PNN) to differentiate different conditions in transformer operation \cite{tripathyrb}. Genetic algorithm-based two parallel Artificial Neural Network (ANN) was used by Balaga et al. to detect and classify faults \cite{balaga}. Bigdeli et al. \cite{bigdeli} proposed Support Vector Machine (SVM) based identification of different transformer winding faults (mechanical and short circuits). Patel et al. \cite{patel} reported that the Relevance Vector Machine (RVM) performs better than PNN and SVM for fault detection and classification. Wavelet-based protection and fault classification in an Indirect Symmetrical Phase Shift transformer (ISPST) was used by Bhasker et al. \cite{tencon} \cite{indicon}. Shin et al. \cite{shin} used a fuzzy-based method to overcome mal-operations and enhance fault detection sensitivities of conventional differential relays. Segatto et al. \cite{segatto} proposed two different subroutines for transformer protection based on ANN. Shah et al. \cite{shah} used Discrete Wavelet Transforms (DWT) and SVM based differential protection. Barbosa et al. \cite{Barbosa} used Clarke’s transformation and Fuzzy logic based technique to generate a trip signal in case of an internal fault. RF classifier was used to discriminate internal faults and inrush in \cite{Shahrfc}.
Ensemble-based learning was used to classify 40 internal faults in the ISPST and the performance was compared with ANN and SVM in \cite{pallav2}. In \cite{systempallav} DT based algorithms were used to discriminate the internal faults and other transient disturbances in an interconnected system with ISPST and power transformers. DWT  and  ANN were used for the detection and classification of internal faults in a two-winding three-phase transformer in \cite{fltcls}. Applicability of seven machine learning algorithms with different sets of features as inputs is used to detect and locate faults in \cite{ietpallav}.

Avoiding mal-operation of differential relays during magnetizing inrush and mis-operation during internal faults by correct detection of faults and inrush ensures the security and dependability of the protection system. Moreover, the classification of the internal faults may provide information about the faulty side of the transformer and may help in the evaluation of the amount of repair and maintenance needed. In a similar direction as in the above-cited publications, this study attempts to use the power of machine learning algorithms to segregate faults from inrush and then determine the type of internal faults in power transformers. The main contributions of this paper are:
\begin{itemize}
    \item 11,088 cases for 11 different internal faults on the primary and secondary sides (1,008 cases for each) and 720 cases of magnetizing inrush are simulated in the transformer.
    \item  A series of features belonging to the time, frequency, and time-frequency domains are extracted. DT distinguishes faults and inrush with an accuracy of 100\% with Sample entropy as feature. Gradient Boost Classifier gives an accuracy of 95.4\% for classifying the internal faults with change quantile and absolute energy as features. 
\end{itemize}

The rest of the paper is organized as follows. Section II illustrates the modeling and simulation of internal faults and magnetizing inrush in PSCAD/EMTDC. Section III describes the extraction and selection of features to differentiate faults from inrush and then classify the faults. Section IV and V consist of the classification framework and results respectively. Section VI concludes the paper. 

\begin{figure}[htp]
\centerline{\includegraphics[width=3.5 in, height=2.0 in]{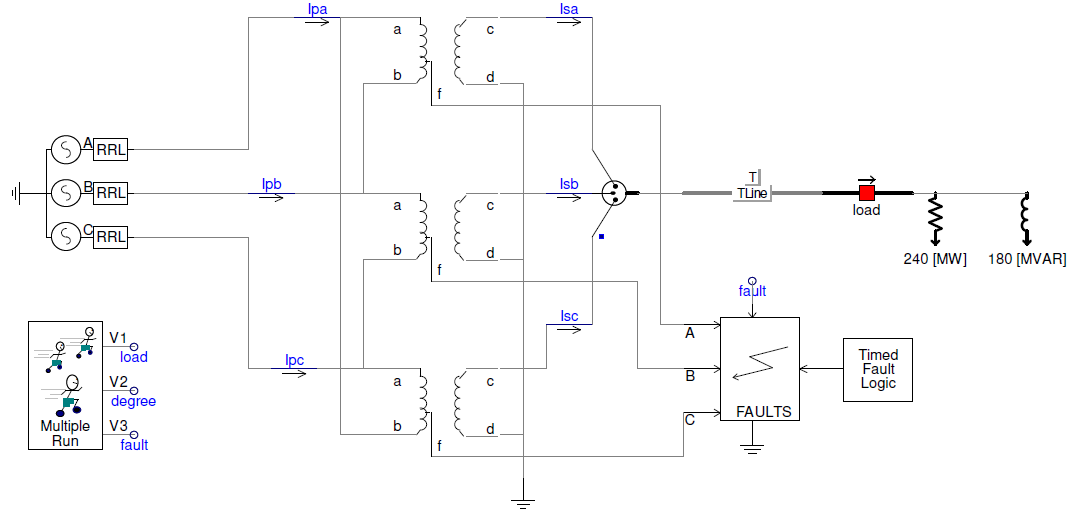}}
\caption{Transformer model showing the ac source, power transformer, multiple-run, faults, transmission line, and load components in PSCAD/EMTDC}
\label{trans}
\end{figure}

\begin{figure*}[htp!]
\centerline
{\includegraphics[width=7.2 in, height=2.9 in]{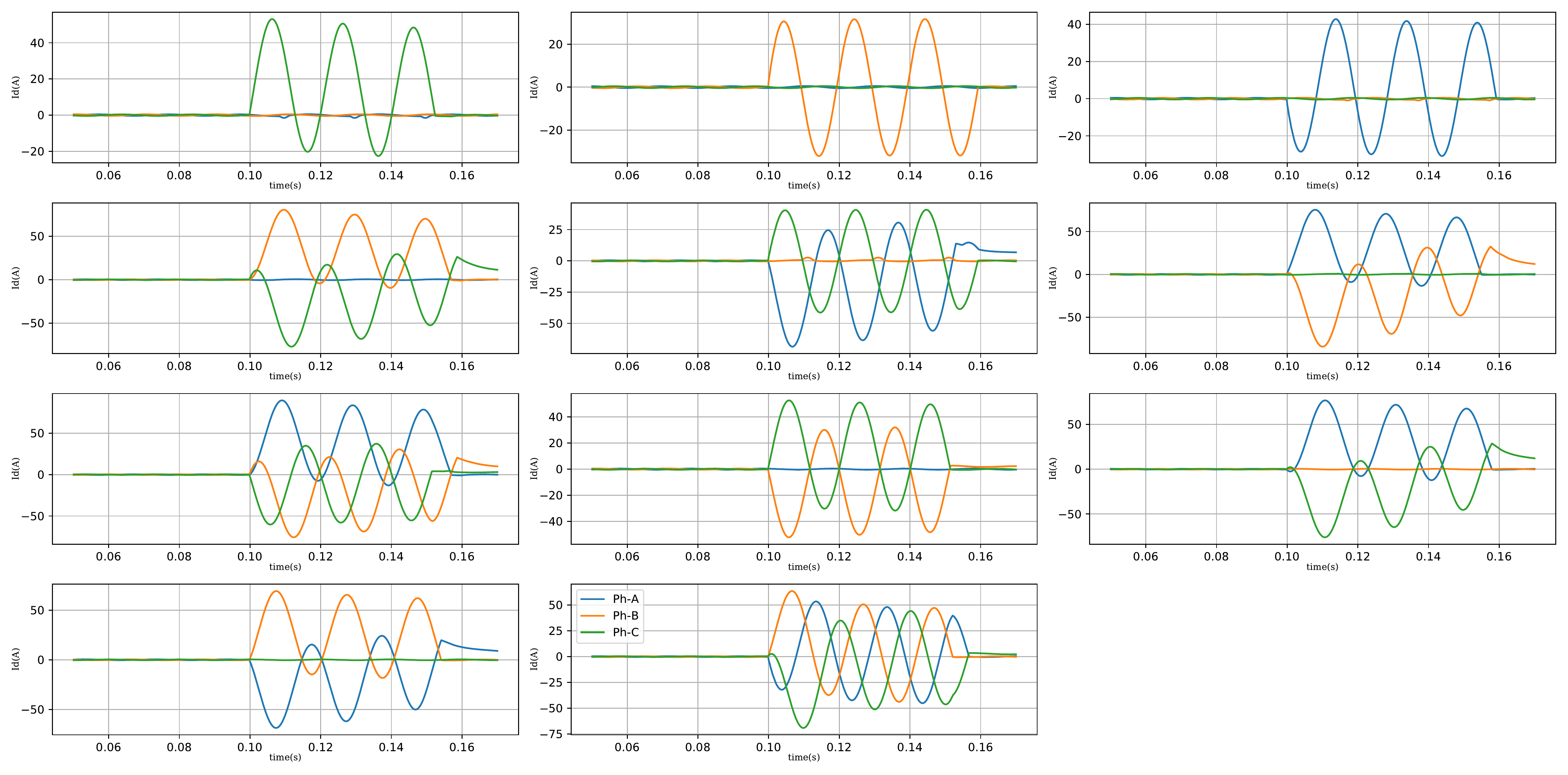}}
\caption{Typical 3-phase differential currents from left to right, top to bottom for (1) a-g, (2) b-g, (3) c-g, (4) ab-g, (5) ac-g, (6) bc-g, (7) 3-phase-g, (8) ab, (9) ac, (10) bc,  and (11) 3-phase internal faults in the primary of the transformer}
\label{Figure_1}
\end{figure*}

\begin{section}{System Modeling}
PSCAD/EMTDC version 4.2 is used for the modeling and simulation of the internal faults and magnetizing inrush in the power transformer. Figure \ref{trans} shows the model consisting of the ac source, transmission line, power transformer, multi-run component, faults component, and a 3-phase load working at 60Hz. The source is rated at 400kV, the transformer is rated at 315 MVA and 400kV/230kV, the transmission line is rated at 230kV with 100km length, and the load is at 240 MW and 180 MVAR. Generally, the operation of a power transformer can be categorized in normal operation, internal fault, external fault, overexcitation, and magnetizing inrush or sympathetic inrush conditions. This paper specifically focuses on the detection and classification of the internal faults in the transformer. 

Winding phase-g faults (a-g, b-g, c-g), winding phase-phase-g faults (ab-g, ac-g, bc-g), winding phase-phase faults (ab, ac, bc), 3-phase and 3-phase-g faults are simulated using the multi-run component. The simulation run-time, fault inception time, and fault duration time are 0.2 secs, 0.1 secs, and 0.05 secs (3 cycles) respectively. The internal faults are simulated on the primary and secondary sides of the transformer. Different parameters of the transformer are varied to get data for training and testing. The fault inception angle is varied from $0\degree$ to $345\degree$  in steps of $15\degree$, the values of fault resistance are: 1, 5, and 10 ohms,  and the percentage of winding under fault is varied from 20\% to 80\% in steps of 10\% in the primary and secondary sides under load and no-load conditions. Consequently, 11,088 cases of differential currents for internal faults are generated with 1,008 cases for each of the 11 different internal faults. Figure \ref{Figure_1} shows the 3-phase differential currents of the 11 different internal faults.

The magnetizing inrush currents are obtained by varying the switching angle from $0\degree$ to $345\degree$ in steps of $15\degree$ with residual flux of $\pm80\%, \pm40\%, 0\% $ in phases a, b, and c for load and no-load conditions. The incoming transformer is connected to the source at 0.1s. As a result, 720 cases of magnetizing inrush are simulated. Figure \ref{inr} shows the 3-phase differential currents of a typical magnetizing inrush condition for a residual flux of 0\%.
\begin{figure}[ht!]
\centering
\includegraphics[width=3.5 in, height= 1.22 in]{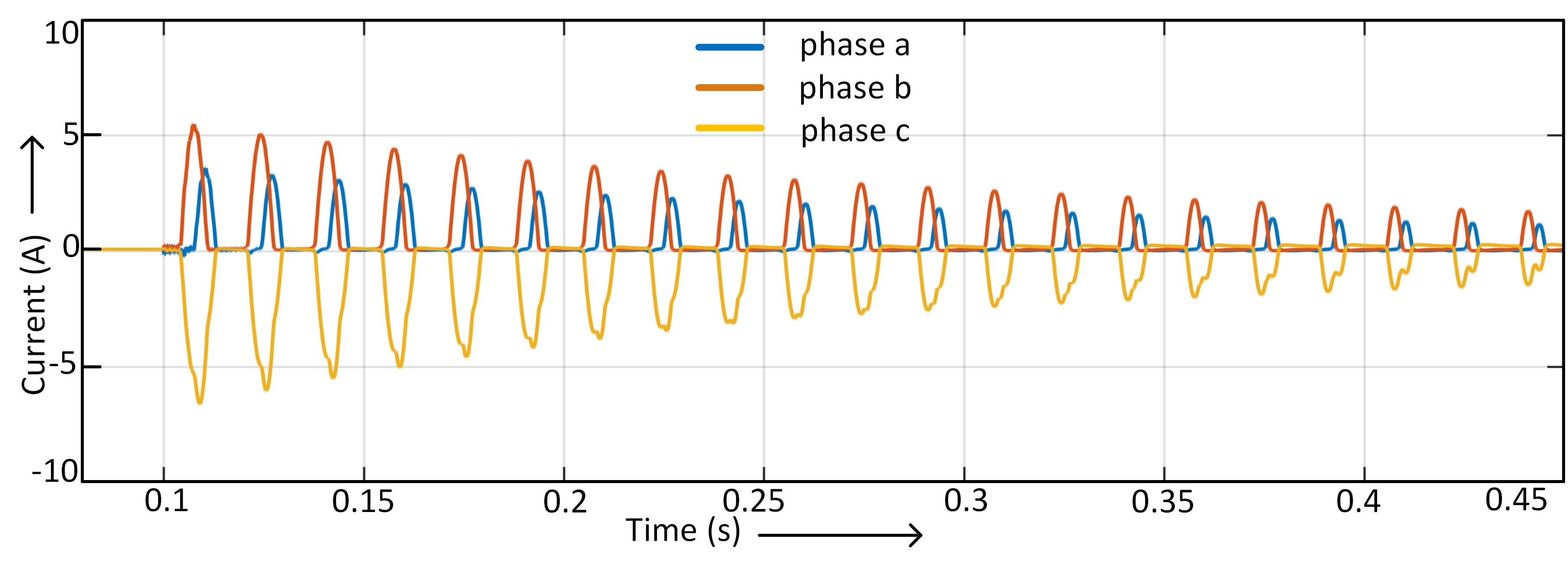}
\caption{3-phase magnetizing inrush differential currents}
\label{inr}
\end{figure}
\end{section}

\begin{section}{Extraction and Selection of Features}
The differential currents used to extract the features are time-series signals which can be differentiated in many ways. The similarity between time series can be established at data-level using Euclidean or Dynamic Time Warping (DTW) \cite{DTW} distance measures. The differential currents from a distinct fault type can also be differentiated from rest using features (e.g., mean, standard deviation (std), frequency, entropy, skewness, kurtosis, or wavelet coefficients) and the distance between the features \cite{TimeSeriesFeatureRep} \cite{NanopoulosTimeSeriesFeat}. Wang et al. \cite{WangTimeSeriesFeat} extracted features from trend, seasonality, periodicity, serial correlation, skewness, kurtosis, chaos, non-linearity, and self-similarity. Wirth et al. \cite{MultiVariateStructure} further extended this approach to multivariate time series signals. Kumar et al. \cite{AbenaPrimo} used a variety of features from time and frequency domains to distinguish binary classes.  

Here, features are extracted from the 3-phase differential currents considering time-domain (e.g., minimum, maximum, median, number of peaks, mean, skewness, number of mean crossings, quantiles, and absolute energy), information-theoretic (sample entropy, approximate entropy, and binned entropy), and coefficients of auto-regression, discrete wavelets, and fast Fourier transforms. A number of features are extracted from the a,b, and c phase differential currents. More information on these features can be found in \cite{tsfresh}. Sample entropy computed from each of the three phases is used to distinguish an internal fault from an inrush. To classify the internal faults three most relevant features are then obtained by comparing the relative importance of the features by using Random Forest in Scikit Learn. The three features are a-phase change quantile, b-phase absolute energy, and a-phase change quantile. Change quantile calculates the average absolute value of consecutive changes of the time series inside two constant values qh and ql. Absolute energy is the sum of the squares of all the data points of the time series. Change quantile and absolute energy are expressed mathematically as given by equation (1) \& equation (2) respectively.

\begin{figure}[ht!]
\centering
\vspace{0.5cm}
\includegraphics[width=3.45 in, height= 1.2 in]{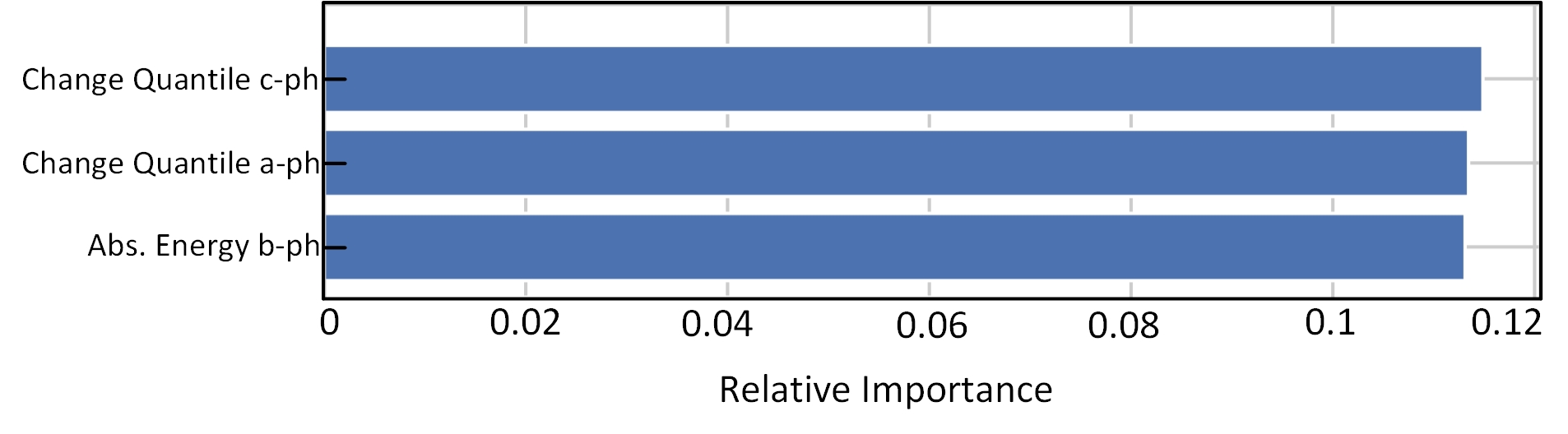}
\caption{Relative importance of the top three features.}
\label{rel}
\end{figure}

\begin{equation} Change\ quantile = \frac{1}{n}\cdot{\sum_{a=1}^{n-1} |S_{a+1} - S_{a}| }
\end{equation}
\begin{equation}Abs.\ energy = \sum_{a=1}^{n} S_a^{2} 
\end{equation}
where, n is the total number of data points in the differential current for phase a, b, and c considered, 
S represents phase a, b, and c differential currents. 

The relative importance of the three selected features are shown in Figure \ref{rel}. These features are used as the input to the classifiers. The importance of a feature (f) at node j is calculated by optimising the objective function ${I(_j,_f)}$  
$$I(_j,_f) = w_j\cdot I(D_p)-\frac{N_l}{N_p}\cdot I(D_l)-\frac{N_r}{N_p}\cdot I(D_r)$$
where, f is the feature to perform the split, $w_j $ = number of samples that reach the node, divided by the total number of samples, $D_p$, $D_l$ and $D_r$ are the dataset of the parent and child nodes, I is "gini" impurity measure, and ${N_p},{N_l}$ and ${N_r}$ are number of samples at the parent and child nodes.
\end{section}

\begin{figure}[ht!]
\centering
\includegraphics[width=3.55 in, height= 1.25 in]{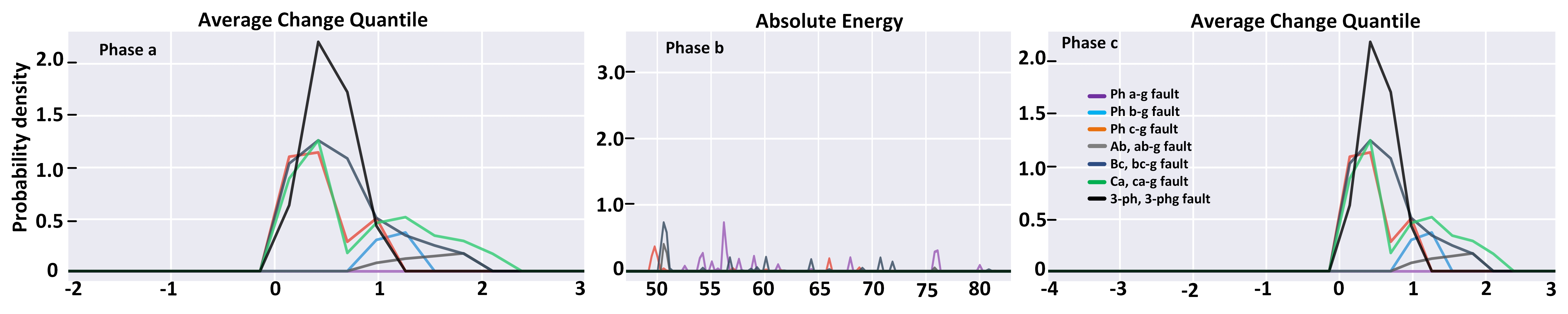}
\caption{Kernel Density Estimate plots showing the probability distribution of the selected features for the seven internal faults}
\label{kdeplot}
\end{figure}

Kernel density estimation plot is a useful statistical tool to picture data shape. The shapes of the probability distribution of multiple continuous attributes for different classes can be visualized in the same plot. In this Parzen–Rosenblatt window method \cite{parzen1} is used to estimate the underlying probability density of the three selected features for the seven different internal faults. Figure \ref{kdeplot} shows the kernel density estimation plots for the chosen features in phase a, b, and c. The Gaussian kernel function is used for approximation of the univariate features with a bandwidth of 0.1 for average change quantile in phases a and c, and bandwidth of 0.001 for absolute energy in phase b.

\section{Fault Detection \& Classification} 
\begin{figure}[htp!]
\centerline{\includegraphics[width=2.4 in, height= 2.9 in]{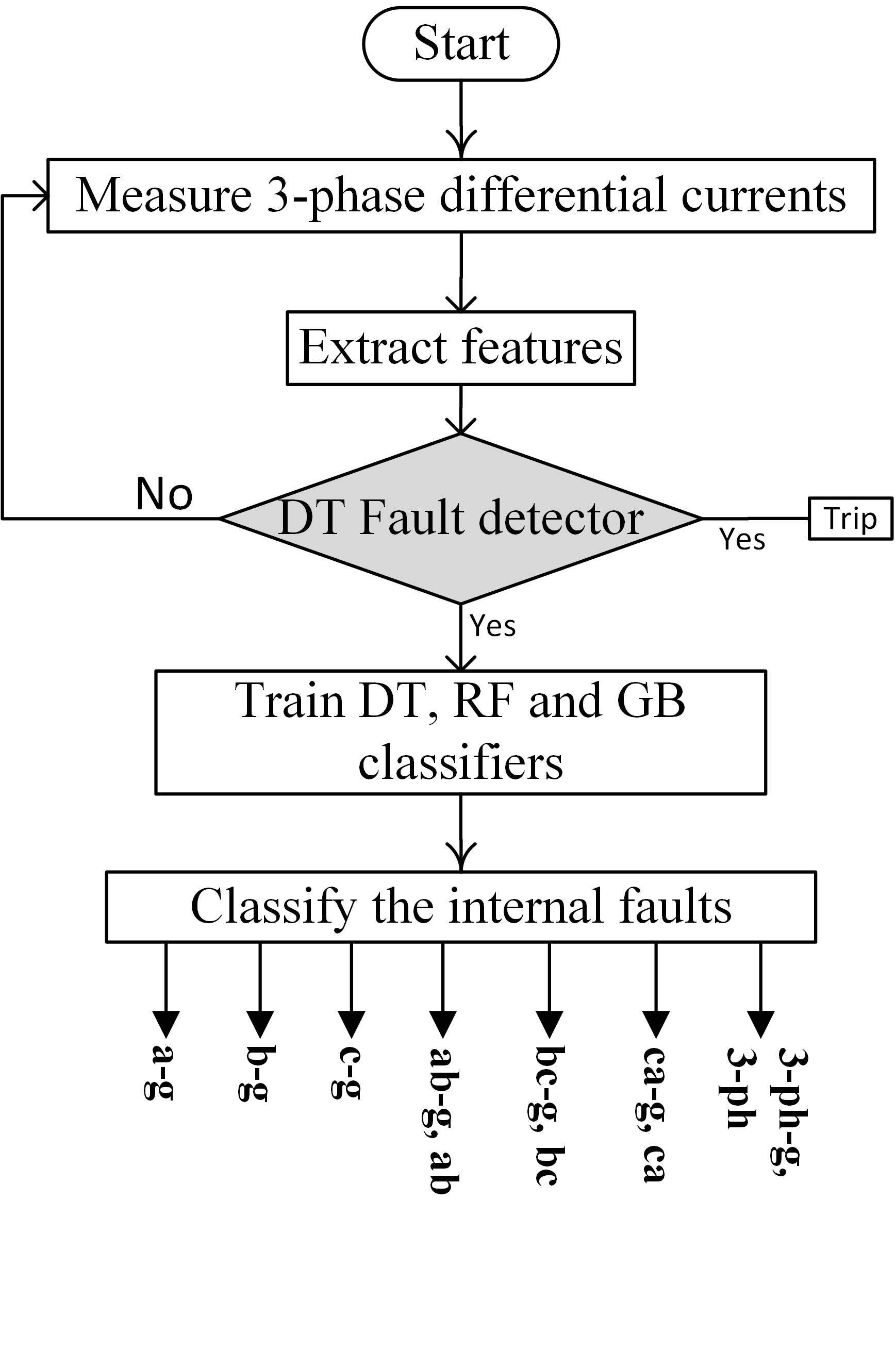}}
\caption{ Fault classification framework.}
\label{flowchart}
\end{figure}
Percentage restraint differential protection compares the operating current and restraining current and thus differentiates external faults and normal operating conditions from the internal faults. Figure \ref{flowchart} shows the detection and classification framework in use. The work in this paper is applicable for internal faults, magnetizing inrush, and normal operation in the power transformer. A DT trained on sample entropy of one-cycle of 3-phase differential currents detects an internal fault or an inrush. It sends a trip signal in case an internal fault is detected and computes change quantile and absolute entropy from the 3-phase differential currents. DT, RFC, and GB are trained on these features to classify the internal faults into seven different types of faults. The training of the classifiers is carried out on 4/5th and testing on 1/5th of the data and grid search is used to find the best hyperparameters.

Three different classifiers are used for training and testing. The first classifier used is the Decision Tree (DT) \cite{dtb} \cite{Quinlan}. DT classifier works on the principle of splitting the data on the basis of one of the available features which gives the largest Information Gain (IG). The splitting is repeated at every node till the child nodes are pure or they belong to the same class. IG is the difference between the impurity of the parent node and the child node. The impurity of the child node decides the (IG). DT is easier to interpret, can be trained quickly and, can model a high degree of nonlinearity in the relationship between the target and the predictor variables \cite{dtadv}.

The second classifier is Random Forest (RF). RF classifiers are a collection of decision trees that use majority vote of all the decision trees to make predictions. The trees are constructed by choosing random samples from the total training sample with replacement \cite{Breiman2001}. The n\_estimators hyperparameter which denotes the number of trees is the important parameter to be tuned. "Grid Search" is used to tune the number of trees in the RF classifier in this case.

The third classifier used is the Gradient Boost (GB) \cite{GB}. GB is also a collection of decision trees like RF. Unlike RF, in GB the trees are added in an iterative manner where each tree learns from the mistakes of previous trees. Thus, the learning rate becomes an important hyperparameter in GB. Higher learning rate and more number of trees increase the complexity of the model.

\begin{table}[htbp!]

\scriptsize
\centering
\caption{Classification Results for Decision Tree}
\setlength{\tabcolsep}{3pt}
\setcellgapes{0 pt}\makegapedcells
\begin{tabular}{|c|c|c|}
\hline
\textbf{\begin{tabular}[c]{@{}c@{}}Actual Fault Type\end{tabular}}&

\textbf{\begin{tabular}[c]{@{}c@{}}Predicted Fault  type\end{tabular}} &

\textbf{\begin{tabular}[c]{@{}c@{}}\# of  misclassified\\ cases\end{tabular}} \\ \hline 

a-g(201) & - & 0  \\ \hline
b-g(218)& ab-g,ab & 25 \\ \hline
c-g(214) & bc-g, bc & 6 \\ \hline

ab-g, ab (398) & b-g  & 22  \\ \hline

bc-g, bc(384) & c-g & 10 \\ \hline

ca-g,ca(411) & 3-ph, 3-ph-g & 37 \\ \hline

3-ph, 3-ph-g(402) & ca-g,ca &42 \\ \hline

\end{tabular}
\label{dttable}

\end{table}

\begin{table}[htbp!]

\scriptsize
\centering
\caption{Classification Results for Random Forest}
\setlength{\tabcolsep}{3pt}
\setcellgapes{0 pt}\makegapedcells
\begin{tabular}{|c|c|c|}
\hline
\textbf{\begin{tabular}[c]{@{}c@{}}Actual Fault Type\end{tabular}}&

\textbf{\begin{tabular}[c]{@{}c@{}}Predicted Fault  type\end{tabular}} &

\textbf{\begin{tabular}[c]{@{}c@{}}\# of  misclassified\\ cases\end{tabular}} \\ \hline 

a-g(184) & - & 0  \\ \hline
b-g(198)& ab-g, ab & 10 \\ \hline
c-g(231) & bc-g, bc & 11 \\ \hline

ab-g, ab (394) & b-g & 11  \\ \hline

bc-g, bc (417) & c-g & 4 \\ \hline

ca-g,ca(388) &3-ph, 3-ph-g & 28 \\ \hline

3-ph, 3-ph-g(406) & ca-g,ca & 45 \\ \hline

\end{tabular}
\label{rftable}

\end{table}

\begin{table}[htbp!]

\scriptsize
\centering
\caption{Classification Results for Gradient Boost}
\setlength{\tabcolsep}{3pt}
\setcellgapes{0 pt}\makegapedcells
\begin{tabular}{|c|c|c|}
\hline
\textbf{\begin{tabular}[c]{@{}c@{}}Actual Fault Type\end{tabular}}&

\textbf{\begin{tabular}[c]{@{}c@{}}Predicted Fault  type\end{tabular}} &

\textbf{\begin{tabular}[c]{@{}c@{}}\# of  misclassified\\ cases\end{tabular}} \\ \hline 

a-g(194) & - & 0  \\ \hline
b-g(190)& ab-g, ab & 11 \\ \hline
c-g(214) & bc-g, bc & 8 \\ \hline

ab-g, ab (413) & b-g & 13  \\ \hline

bc-g, bc (381) & c-g & 4 \\ \hline

ca-g,ca(404) &3-ph, 3-ph-g & 35 \\ \hline

3-ph, 3-ph-g(414) & ca-g,ca &31 \\ \hline

\end{tabular}
\label{svmtable}

\end{table}

\section{Results}
\subsection{Fault detection}
Sample entropy computed from one-cycle (167 samples) of 3-phase differential currents is used to detect an internal fault or an inrush. DT distinguishes internal faults from magnetizing inrush with 100\% accuracy. Since the number of cases for faults(11088) and inrush (720) is not balanced Synthetic Minority Over-Sampling  Technique (SMOTE) \cite{SMOTE} is used to create minority synthetic data and NearMiss algorithm is used for under-sampling the majority class.

\subsection{Fault classification}
At first, attempts were made to classify the internal faults into 11 classes. But, it was observed that line to line to ground faults were misclassified as line to line faults and 3-phase faults as 3-phase-g. So, the line to line faults and line to line to ground faults are merged in one class. For instance fault types ab and ab-g form one class. Similarly, 3-phase and 3 phase-g faults form one class. After merging these internal fault types, the resultant number of classes became 7. The first three classes a-g, b-g, and c-g consist of 1008 samples, and the rest of the classes consist of 2016 samples. 
The misclassification for DT, RF and GB classifiers between different types of internal faults are reported in Table \ref{dttable},  Table \ref{rftable}, and Table \ref{svmtable} respectively.

The hyperparameters used for the Decision Tree Classifier are: criterion = 'gini', min\_samples\_leaf = 1, and  min\_samples\_split = 2. The training and testing accuracies obtained are 93.65\% and 93.6\% respectively.
The best hyperparameters obtained using "grid search" for the Random Forest Classifier are: criterion = 'gini', n\_estimators = 595, min\_samples\_leaf = 1, and min\_samples\_split = 2. The training and testing accuracies obtained in this case are 93.61\% and 95.1\% respectively. For Gradient Boost Classifier the best hyperparameters obtained using "grid search"  are: learning\_rate = 0.1, max\_depth = 10, n\_estimators = 10000, criterion = friedman\_mse, min\_samples\_leaf = 1, and min\_samples\_split = 2. The training and testing accuracies obtained for GB are 93.95\% and 95.4\% respectively. The default hyperparameter values are used for the rest of the hyperparameters for all three classifiers.

Pre-processing, extraction of features, and selection of important features are done in Python 3.7 and MATLAB 2019, and the DT, RFC, and GB classifiers are built in Python 3.7 using Scikit-learn machine learning library \cite{scikit}. Intel Core i7-6560U CPU @ 2.20 GHz having 8 GB RAM is used to perform the simulations and for training the classifiers.

\section{Conclusion}
 Firstly, magnetizing inrush is distinguished from the internal faults in a power transformer in one cycle with DT and secondly, the three DT based classifiers are trained to classify the internal faults by using the most relevant features obtained from the differential currents in phases a, b, and c in this study. The random forest feature selection was used to reduce the number of features. The DT distinguishes faults from inrush currents without a single error. The GB classifier achieved the best accuracy of 95.4\% for fault classification whereas, the DT has the lowest accuracy of 93.6\% among the three classifiers. In this paper, only the internal faults and magnetizing inrush in the power transformer were considered. In the future, over-excitation, sympathetic inrush, turn-to-turn faults, and inter-winding faults can be considered for detailed analysis.

\balance
\bibliographystyle{IEEEtran}
\bibliography{references} 
\end{document}